\documentclass{elsart5p}

\usepackage{graphicx}

\usepackage{amssymb}

\begin{document}

\begin{frontmatter}



\title{Site-Selective NMR in the Quasi-1D Conductor $\beta$-Sr$_{0.33}$V$_{2}$O$_{5}$}
\author{T. Waki\corauthref{cor1}},
\corauth[cor1]{Corresponding author. tel/fax: 81-4-7136-3228}
\ead{twac@issp.u-tokyo.ac.jp}
\author{M. Takigawa},
\author{T. Yamauchi},
\author{J. Yamaura},
\author{H. Ueda},
\author{Y. Ueda}

\address{Institute for Solid State Physics, University of Tokyo,
Kashiwanoha, Kashiwa, 277-8581, Japan}

\begin{abstract}
We report $^{51}$V NMR experiments in the metallic phase of the quasi-one-dimensional 
(1D) conductor $\beta$-Sr$_{0.33}$V$_{2}$O$_{5}$.  The Knight shift and the
quadrupole splitting of all six vanadium sites were determined as a function of the
direction of magnetic field perpendicular to the conducting $b$-axis. Magnetic
properties of the three 1D structural units are remarkably heterogeneous.  In
particular, the V2 ladder unit shows pronounced charge disproportionation among the
two (V2a and V2b) sites. At one of these sites, the absolute value of the Knight shift and
$1/(T_{1}T)$ increase steeply with decreasing temperature, suggesting development
of ferromagnetic correlation.      
\end{abstract}

\begin{keyword}
A. oxides \sep D. magnetic properties \sep D. nuclear magnetic resonance (NMR)

\PACS 71.30.+k \sep 76.60.-k
\end{keyword}
\end{frontmatter}

\section{Introduction}
The family of quasi-one-dimensional mixed valent vanadium oxides
$\beta$-$A_{0.33}$V$_{2}$O$_{5}$ ($A$=Li, Na, Ag, Ca, Sr, and Pb) exhibit a
variety of exotic phenomena such as metal-insulator transition, charge ordering,
and pressure-induced superconductivity~\cite{Ueda,Yamauchi}. They provide
fascinating opportunities to study mechanism for competing orders in a common 
monoclinic crystal structure shown in Fig.~1~\cite{Sellier}.  There are three types
of one-dimensional V-O network extending along the most conducting $b$-axis: a zigzag
chain formed by the V1 sites, another zigzag chain formed by the V3 sites, and a
two-leg ladder formed by the V2 sites.  At high temperatures, the $A$ ions are
distributed randomly over two sites with 50~\% occupancy in the space group
$C2/m$.  They order below a certain temperature and fully occupy a single sites
in the new space group $P2_{1}/a$.  This doubles the unit cell along the $b$-axis
and generates two inequivalent sites for each of the three vanadium sites, which
are denoted as Vna and Vnb (n=1 - 3, see Fig.~1).

\begin{figure}[b]
 \includegraphics[width=1\linewidth]{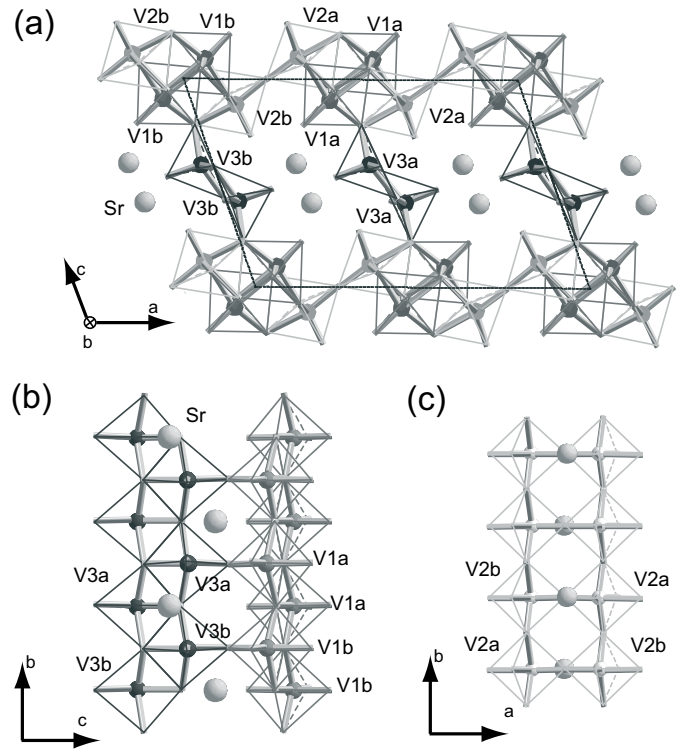}
 \caption{Crystal structure of $\beta$-Sr$_{0.33}$V$_{2}$O$_{5}$ with the space 
group $P2_{1}/a$ viewed from the (a) $b$-, (b) $a^{\ast}$-, and (c) $c$-directions. }
 \label{fig-1}
\end{figure}
This ordering occurs above the room temperature (RM) in 
$\beta$-Sr$_{0.33}$V$_{2}$O$_{5}$, which is a Pauli paramagnetic metal at RT.  
It undergoes a metal-insulator transition at $T_{\mathrm{MIT}}\sim$ 170 K 
accompanied by a simultaneous structural transition, that further triples the 
unit cell along the $b$-axis \cite{Sellier, Yamaura}.  The magnetic susceptibility in 
the low $T$ insulating phase indicates a singlet ground state with a spin gap.  
A spatial order of vanadium valence into V$^{4+}$ ($3d^{1}$) and V$^{5+}$ 
($3d^{0}$) with 1:2 ratio was suggested for the low $T$ phase from structural 
analysis.  However, the neutron diffraction~\cite{Nagai} and the nuclear magnetic resonance 
(NMR)~\cite{Itoh} experiments ruled out such atomic localization of $d$ electrons for   
$\beta$-Na$_{0.33}$V$_{2}$O$_{5}$, which also shows simultaneous metal-insulator 
and structural transitions.  In this paper we report results of $^{51}$V-NMR 
experiments on a single crystal of $\beta$-Sr$_{0.33}$V$_{2}$O$_{5}$ and 
discuss charge distribution and spin dynamics in the high $T$ metallic phase.  
NMR data on $\beta$-Sr$_{0.33}$V$_{2}$O$_{5}$ have been reported 
only for a powder sample so far~\cite{Yasuoka}.   

\section{Experiment}
A single crystal of $\beta$-Sr$_{0.33}$V$_{2}$O$_{5}$ obtained by the Czochralski 
method was used in this work.  It has the size 0.5$\times$1.8$\times$0.2~mm$^{3}$
with the largest dimension along the $b$-axis. The $^{51}$V NMR 
spectra were obtained by summing the Fourier transform of the spin-echo signals 
with the frequency step of 0.3 MHz.  The nuclear spin-lattice relaxation rate 
1/$T_{1}$ was measured by the inversion recovery method.  The magnetic field 
$H_{0}$ (10.509~T) was applied in the $ac$-plane.  The direction of the field is 
specified by the angle ($\theta$) measured from the $a$-axis. 

\section{Results and Discussions}
There are four V atoms for each of the six crystalographic sites in a unit 
cell of the metallic phase (space group $P2_{1}/a$).  When the magnetic 
field is applied in the $ac$-plane, these four atoms become equivalent and give 
identical NMR spectra.  Since $^{51}$V nuclei have spin $I$=7/2, the electric field 
gradient (EFG) splits the NMR spectrum into equally spaced seven lines at the 
following frequencies due to first order effect of the nuclear quadrupole
interaction,    
\begin{equation}
\nu_{m}(\theta)=\{1+K(\theta)\}\gamma H_{0}+(m-1/2)\nu_{Q}(\theta) \ ,
\end{equation}
where $\gamma$ is the nuclear gyromagnetic ratio (11.193~MHz/T), 
$K(\theta)$ is the Knight shift, $\nu_{Q}(\theta)$ is the quadrupole splitting, 
and $m$ specifies the nuclear transitions $I_{z}=m \leftrightarrow m-1$.  

\begin{figure}[t]
 \includegraphics[width=1\linewidth]{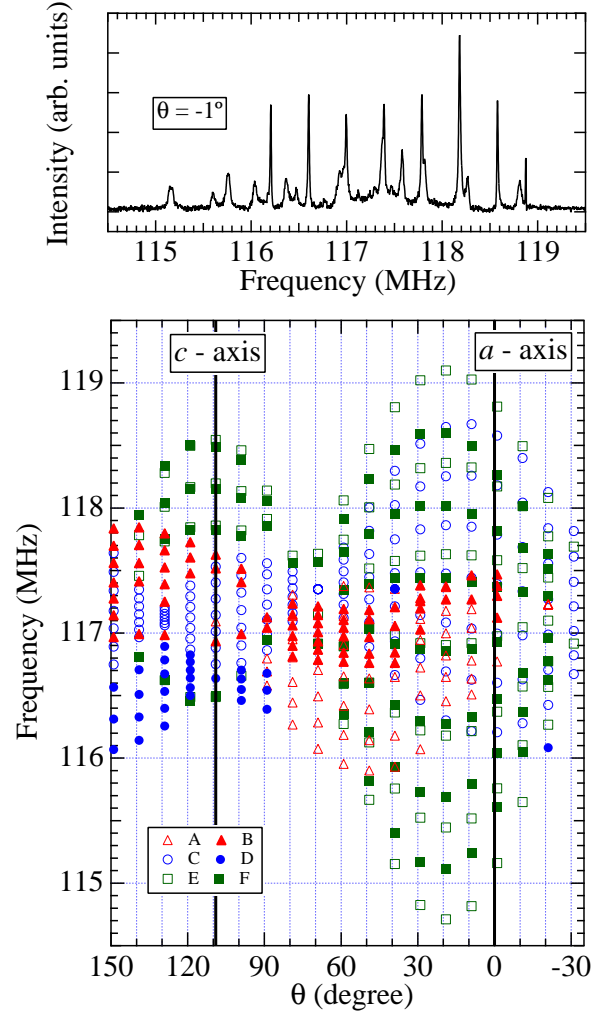}
 \caption{An example of the NMR spectrum at (upper panel) and the  
angle dependence of the NMR frequency of all the observed lines (lower panel)
at $T$=190~K.}
 \label{fig-2}
\end{figure}
The upper panel of Fig.~\ref{fig-2} shows the NMR spectrum at $\theta=-1^{\circ}$,
while the lower panel displays the angle dependence of the resonance frequencies for
all the lines at $T$=190~K. At each angle, there are several sets of equally spaced
lines assigned as A - F.  In many cases only a portion of the quadrupole split seven
lines are clearly resolved.  However, they are sufficient to determine
$K(\theta)$ and $\nu_{Q}(\theta)$ from Eq.~(1).
\begin{figure*}[t]
 \includegraphics[width=1\linewidth]{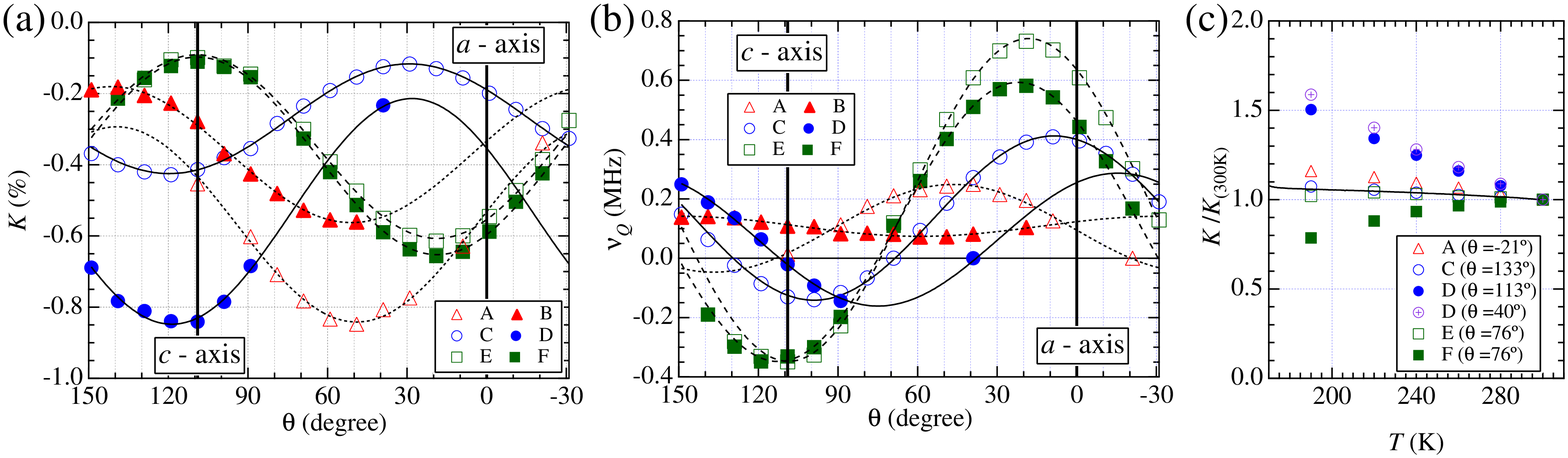}
 \caption{Angle dependence of (a) $K$ and (b) $\nu_{Q}$ for all sites at $T$=190~K.  The solid lines 
show the fitting to Eq. (2).  (c) $T$-dependences of $K$ normalized by the value at 300~K 
(symbols) and the bulk susceptibility (the solid line).}
 \label{fig-3}
\end{figure*}
In Fig.~3(a) and (b), we plot the angle dependences of $K(\theta)$ and
$\nu_{Q}(\theta)$ for the A - F sites.  Since $K(\theta) \ll 1$ and $\nu_{Q}(\theta) \ll \gamma H_{0}$, 
we need consider only the lowest order effect for the frequency shift. Then
$K(\theta)$ and $\nu_{Q}(\theta)$ should obey the following angle
dependences~\cite{Slichter}, 
\begin{eqnarray}
K(\theta) & = & K_{1}+K_{2} \cos^{2}(\theta-\psi_{K}) \nonumber \\ 
\nu_{Q}(\theta) & = & \nu_{1}+\nu_{2} \cos^{2}(\theta-\psi_{Q})  \ . 
\end{eqnarray}
These equations indeed fit the data very well as shown by the solid, dashed and dotted lines 
in Fig.~3(a) and (b).  Thus we have completely determined $K(\theta)$ and 
$\nu(\theta)$ for all the vanadium sites. 

Generally the Knight shift at transition metal sites consists of the orbital and the spin 
contributions, $K(\theta)=K_\mathrm{orb}(\theta)+K_\mathrm{spin}(\theta)$.  
While $K_\mathrm{spin}$ may be either negative or positive, 
$K_\mathrm{orb}$ is always positive.  The fact that $K(\theta)$ is  
negative for the entire range of angle at all sites indicates that the dominant
term is $K_\mathrm{spin}$, which is related to the local spin susceptibility as 
$K_\mathrm{spin}(\theta)=\{A_\mathrm{cp}+A_\mathrm{dip}(\theta)\}\chi_\mathrm{spin}$.  
The core polarization coupling constant $A_\mathrm{cp}$ is isotropic and negative, while  
the dipolar coupling constant $A_\mathrm{dip}(\theta)$ is purely anisotropic, i.e. 
its directional average is zero.  Since $\chi_\mathrm{spin}$ is nearly isotropic 
except for small anisotropy of the $g$-value,  the observed angular variation 
of $K(\theta)$ must be ascribed to the anisotropy of $A_\mathrm{dip}(\theta)$, which is 
determined by the spin density distribution of $d$-electrons.  

In $\beta$-Sr$_{0.33}$V$_{2}$O$_{5}$, each vanadium atom is surrounded  
by a distorted squared pyramid formed by oxygen atoms.  The shortest vanadyl bonding 
between the vanadium and the apical oxygen lies nearly in the $ac$-plane 
for all the V sites (within 4$^{\circ}$).  Doublet and Lepetit investigated the local electronic 
structure based on the extended H\"{u}ckel tight-binding calculation~\cite{Doublet}.  
They proposed that the magnetic orbital accommodating the $d$-electrons is the 
$d_{xy}$-orbital, where the local $z$-axis is parallel to the shortest V-O bonding 
and $x$ and $y$ point toward the oxygen atoms in the basal plane of the pyramid.  
The dipolar coupling $A_\mathrm{dip}$ to the spin density on the $d_{xy}$-orbital 
takes the largest negative value when the field is along the $z$-axis.  Since 
$A_\mathrm{cp}$ is negative, $K(\theta)$ is expected to be most negative when 
the field is along the shortest V-O bond direction. 

\begin{table}[b]
\caption{The upper part lists the values of $\psi_{K}$ for the A to F sites.
The lower part shows the directions of the shortest V-O vanadyl bonding projected onto $ac$ plane.
The angles were measured from the $a$-axis.}
\begin{center}
\begin{tabular}{p{50pt} p{30pt} p{30pt} p{30pt} p{30pt} p{30pt} p{30pt}}
\hline\hline
Sites & A & B & C & D & E & F \\
\hline
$\psi_{K}$ & 49.5$^{\circ }$ & 51.3$^{\circ }$ & 120.1$^{\circ }$ & 119.2$^{\circ }$ 
& 19.7$^{\circ }$ & 19.7$^{\circ }$ \\
\hline\hline
Sites & V1a & V1b & V2a & V2b & V3a & V3b \\
\hline
V-O bond & 43$^{\circ }$ & 45$^{\circ }$ & 110$^{\circ }$ & 117$^{\circ }$ 
& 22$^{\circ }$ & 22$^{\circ }$ \\
\hline
\end{tabular}
\end{center}
\label{table_1}
\end{table}
In table \ref{table_1} we compare the values of $\psi_{K}$ at which $K(\theta)$ becomes 
most negative (upper panel) with the directions of the shortest V-O bonding projected 
onto the $ac$-plane obtained from the structural data in Ref.~\cite{Sellier} (lower part).  
We can immediately assign the three groups of sites (A, B), (C, D) and (E, F) to the 
(V1a, V1b), (V2a, V2b) and (V3a, V3b) sites, respectively.  We cannot determine, 
however,  which one in each group corresponds to "a" or "b".  With this assignment 
the directions for the minimum of $K(\theta)$ and the shortest V-O bonding agree within 
ten degrees, supporting that the magnetic orbital indeed has $d_{xy}$ symmetry.  

The data in Fig.~3(a) indicate that while $K(\theta)$ for the V3a and V3b sites are almost 
identical, it differs substantially between the V2a and V2b sites and between the V1a and 
the V1b sites. In particular, the amplitudes of the angular variation of $K(\theta)$ 
($K_{2}$ in Eq.~2) for the V2a and the V2b sites differ by a factor two.  Since the 
angular variation of $A_\mathrm{dip}(\theta)$ should be similar for all the sites, the local spin 
susceptibility $\chi_\mathrm{spin}$ must be significantly different for the V2a and the V2b sites.
Such alternation of $\chi_\mathrm{spin}$ should be ascribed to charge disproportionation.  
In other words, although the $d$-electron density is quite uniform in the V3 zigzag-chains, 
it alternates strongly in the V2 ladders.  The strong charge disproportionation even 
in the metallic phase may appear surprising.  However, this has been predicted 
by Doublet and Lepetit.  They argued that ordering of Sr leads to a large difference 
in the energy of the $d$-levels, therefore, the different charge density at the V2a and 
the V2b sites~\cite{Doublet}.  On the other hand, the three groups of sites, V1, V2, and 
V3, show quite similar values of $K_{2}$ when averaged over the "a" and the "b" sites.  
Thus the $d$-electrons are distributed rather uniformly over the three 1D structural units.   
 
The V2a and the V2b sites exhibit also distinct $T$-dependence of the Knight shift. 
Figure 3(c) show the $T$-dependence of $K$ for several sites normalized by the value at 
$T$=300~K. They were measured at "magic angles" at which $\nu_{Q}(\theta)$=0. (No
measurements were done for the B sites which do not have a magic angle in the $ab$-plane.)
The steep increase of $|K|$ at the D sites with decreasing temperature is in strong contrast 
to the nearly constant behavior at the C sites.  We also note peculiar decrease of $|K|$ 
with decreasing temperature at the F sites.  Although the bulk susceptibility shows little 
$T$-dependence (the solid line), this must be a results of cancellation of opposite 
$T$-dependence at the D and F sites.  We should mention that such strong heterogeneity within 
the same structural unit has not been observed in $\beta$-Na$_{0.33}$V$_{2}$O$_{5}$~\cite{Suzuki}   

The angle dependence of $\nu_{Q}(\theta)$ shown in Fig.~3(b) also reflects the 
local symmetry of each V sites.  There is indeed noticeable correlation
between the directions of shortest V-O bonding and extrema of $\nu_{Q}(\theta)$ 
for many sites. However, EFG is rather difficult to analyze because of contributions 
not only from the on-site $d$-electrons but from the neighboring ions.  Hence we did not 
attempt to make any quantitative analysis.  

\begin{figure*}[t]
\begin{center}
 \includegraphics[width=1\linewidth]{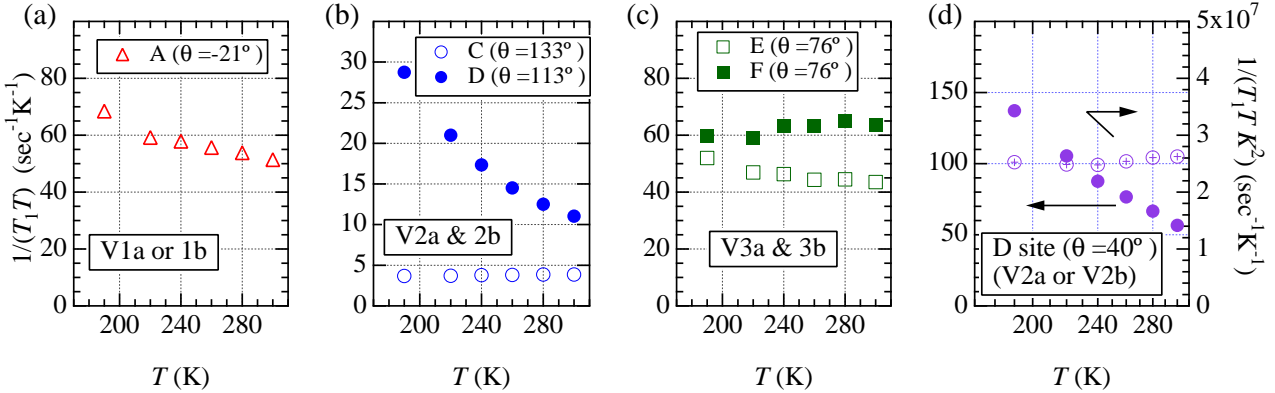}
\end{center}
 \caption{$T$ dependence of $1/T_{1}T$ at various sites measured with the field along 
the magic angle indicated in each panel. Also shown in (d) is $1/(T_{1}TK^{2})$ at the 
D sites.}
 \label{fig-4}
\end{figure*}
We now discuss the spin dynamics. The nuclear relaxation rate 1/$T_{1}$ 
was measured also with the field along magic angles.  This enabled site-selective 
measurements of 1/$T_{1}$ even when the line is not well isolated from others, due to 
strongly enhanced signal, distinct rf-pulse width for optimum excitation, and simple 
exponential time dependence of the nuclear magnetization recovery at magic angles.  
For the E and F sites, which have nearly the same magic angle, the nuclear recovery 
curve was fit to sum of two exponential functions.  The result was checked by separate 
measurements on well isolated quadrupole satellite lines.   

The $T$-dependences of  $1/(T_{1}T)$ are shown in Fig.\ref{fig-4}.  
Again different sites show remarkably contrasting behavior.  The V3a and V3b
sites (E and F) show nearly constant behavior with the similar magnitude of $1/(T_{1}T)$, 
suggesting absence of strong spin correlations.  Precisely speaking, however, 
the E sites show slight increase of $1/(T_{1}T)$ with decreasing temperature.  Such 
increase is more clearly visible for the A sites (one of the V1 sites).  The results for 
the V2a and V2b (C and D) sites are most anomalous.  While $1/(T_{1}T)$ at the 
C sites is nearly constant, it increases strongly at the D sites with decreasing temperature.  
At the D sites, $1/(T_{1}T)$ and $K^2$ follow approximately the same $T$-dependence 
as indicated by nearly constant $1/(T_{1}TK^2)$ shown in Fig. 4(d).  

Generally for quasi 1D electron systems, spin fluctuations both at $q\sim0$ and at 
$q \sim 2k_{F}$ contribute to $1/(T_{1}T)$,     
\begin{equation}
1/(T_{1}T)=1/(T_{1}T)_{q \sim 0}+1/(T_{1}T)_{q \sim 2k_{F}} \ .
\end{equation}
In the Luttinger-liquid (LL) theories, which is expected to be applicable to quasi 
1D conductors, the $T$-dependence of each term is given as 
$1/(T_{1}T)_{q \sim 0}\propto\chi_{spin}^{2}$, $1/(T_{1}T)_{q \sim 2k_{F}}\propto T^{K_{\rho}-1}$, 
where $K_{\rho}$=1 for non-interacting electrons but generally $0<K_{\rho}<1$ for repulsive 
interactions~\cite{Zavidonov,Giamarchi}.  The nearly constant behavior of $1/(T_{1}TK^{2})$ 
(Fig. 4(d)) suggests that the dominant contribution to $1/(T_{1}T)$ at the D sites comes 
from $q\sim0$ spin fluctuations.  Thus our results appear to support strong ferromagnetic 
correlation in the V2 ladder units in the high $T$ metallic phase.  Another peculiar
result is the nearly constant behavior of $1/(T_{1}T)$ at the F sites, in spite of the substantial 
decrease of the Knight shift with decreasing temperature.   

\section{Conclusion}
We performed site-selective $^{51}$V NMR measurements on a single crystal of 
$\beta$-Sr$_{0.33}$V$_{2}$O$_{5}$ in the high $T$ metallic phase.  We found that 
the magnetic properties are remarkably heterogeneous even within the same structural unit. 
In particular, the V2 ladder unit shows strong charge disproportionation caused by the Sr ordering.  
At one of the V2 sites, the absolute value of the Knight shift and $1/(T_{1}T)$ increases strongly 
with decreasing temperature, indicating development of ferromagnetic correlation.  The local spin
susceptibility at one of the V3 sites decreases substantially with decreasing temperature, although
$1/(T_{1}T)$ is nearly constant. 
    
This work was supported by Grant-in Aids for Scientific Research (No. 18104008) from the 
Japan Society for the Promotion of Science, for Scientific Research on Priority Areas on 
"Anomalous Quantum Materials", and for COE Research (No. 12CE2004) from the MEXT Japan.  



\end{document}